\documentclass[prc,aps,showkeys,showpacs,nofootinbib]{revtex4}

\usepackage{graphicx}

\begin{document}
\date{\today}
\title{Stochastic mean-field dynamics for fermions in the weak coupling limit
}

\author{Denis Lacroix}
\address{Laboratoire de Physique Corpusculaire,
ENSICAEN and Universit\'e de Caen,IN2P3-CNRS,
6 Blvd du Mar\'{e}chal Juin,
14050 Caen, France}

\begin{abstract}
Assuming that the  effect of the residual interaction beyond mean-field is weak and
has a short memory time, two approximate treatments of correlation in fermionic
systems by means of Markovian quantum jump are presented.
A simplified scenario for the introduction of fluctuations beyond mean-field is first
presented.  In this theory, part of the quantum correlations between the residual interaction and
the one-body density matrix are neglected and jumps occur between many-body densities formed of pairs of states
$D=\left| \Phi_a \right> \left< \Phi_b \right|/\left< \Phi_b \left.  \right|\Phi_a \right>$  
where $\left| \Phi_a \right>$ and $\left| \Phi_b \right>$ are  
antisymmetrized products of single-particle states. The underlying Stochastic
Mean-Field (SMF) theory is discussed and applied to the monopole vibration of a spherical $^{40}$Ca
nucleus under the influence of a statistical ensemble of two-body contact interaction.
This framework is however too simplistic to account for both fluctuation and dissipation.
In the second part of this work, an alternative quantum jump method is  
obtained without making the approximation on quantum correlations. Restricting to two particles-two holes
residual interaction, the evolution of the one-body density matrix of a correlated system is transformed into
a Lindblad equation. The associated dissipative dynamics can be simulated by quantum jumps
between densities written as $D = \left| \Phi  \right>\left< \Phi  \right|$
where $\left| \Phi \right>$ is a normalized Slater determinant. The associated stochastic Schroedinger equation
for single-particle wave-functions is given.

\end{abstract}


\maketitle

\section{Introduction}
The description of quantum self-interacting systems with many degrees of freedom is common to
many fields of physics, including Bose-Einstein condensates, atomic clusters and nuclear systems.
A striking aspect related to this problem is the emergence of well ordered motion at the same time
as complexity and chaos \cite{Guh98,Zel96}.
In many situations, the self-consistent mean-field theory provides a suitable framework
to describe ordered motions. It also corresponds
to one of the most useful method
to study the static and dynamical properties of self-interacting systems\cite{Rin80,Bla86,Neg88}.
However, it often turns out that mean-field reproduces average properties of one-body observables
but underestimates dissipative and fluctuating aspects. This can directly be assigned to the absence of two-body
effects beyond mean-field. In the nuclear context, it has been proposed to extend mean-field theory by considering that
one-body degrees of freedom represents a subsystem which is coupled to more complex internal degrees of freedom.
By doing so, the problem of self-interacting system has been mapped to an open quantum system problem. In nuclei, extensive
work has been devoted to the formal derivation of dissipative quantum mechanics \cite{Kad62}
and/or related stochastic equations for fermions,
including Markovian and non-Markovian effects \cite{Ayi80,Gra81,Bal81,Ayi88,Ayi01,Rei92,Rei92-2,Oni95}.
These approaches have in common that the
residual part of the interaction introduces disorder on top of the mean field.
These theories end with
rather complex transport equations which are hardly applicable in realistic situations \cite{Lac98-2}.
In fact, only recently, the theory proposed in ref. \cite{Ayi01} has been applied to small amplitude
collective vibrations \cite{Lac01}. Its application to large amplitude motion in non-equilibrated
quantum many-body dynamics remains an open issue \cite{Lac04,Lac98-2}.

During the past decades, large theoretical efforts have been devoted
to the development of Monte-Carlo methods to describe the static properties of many-body
interacting systems \cite{Cep95}. Recent applications to nuclear physics have
shown that stochastic methods can successfully be applied  to
describe the structure of nuclei \cite{Koo97}. These methods can also be extended to dynamical
problems \cite{Neg88} and has been used recently to treat the exact dynamics of interacting bosons  
\cite{Car01} or fermions \cite{Jul02} in schematic cases.
As underlined in ref. \cite{Lac05},
it is highly desirable to provide approximate theory in order to describe dissipation in many-body systems  taking
advantage of recent advances in Monte-Carlo methods. The present work is an exploratory study devoted to the
description of dissipation and fluctuations in dynamical problems using quantum jump techniques.
Starting from a perturbative treatment of the residual interaction, two strategies are used to transform
the many-body dynamics into a stochastic process.
In the first strategy, neglecting part of the quantum correlations, a stochastic theory that may be economical in
terms of numerical implementations is obtained. While appropriate
to treat fluctuations beyond mean-field, it is however not suitable for dissipation. In the second part of this
article, we show that the approximate treatment of quantum correlations can be avoided leading to a more general
framework. In that case, the dynamical equation of motion of one-body degrees of freedom
can be mapped into a Lindblad equation \cite{Lin75,Lin76} generally found in the theory of open
quantum systems \cite{Bre02}. The associated jump process is finally discussed.

\subsection{Generalities on perturbation theory and stochastic mechanics}

We consider a many-body fermionic system described by a two-body Hamiltonian $H$. We assume that the
mean-field theory already provides a good approximation of its static and dynamical
properties. In this case, the N-body wave function can be replaced by an antisymmetrized
product of single particle states interacting through an effective self-consistent mean-field, denoted
by $H_{MF}$. Let us assume that the
system is initially a Slater determinant, denoted by $\left| \Phi (t) \right>$ and let us introduce the mean-field propagator $U_{MF}(t',t)$.
The great advantage of mean-field theory, is that the propagated many-body state $\left| \Phi (t') \right>
= U_{MF}(t',t)\left| \Phi (t) \right>$ remains a Slater determinant and the dynamical evolution
of the system reduces to the evolution of its single-particle components.
Thus, the many-body density $D$ is approximated by $D \simeq
\left| \Phi (t') \right> \left< \Phi (t') \right|$. Accordingly, all the information on the
system is contained in the one-body density matrix, denoted by $\rho$,
whose matrix elements
are defined by
$\left< j \left| \rho \right| i\right> = Tr\left( a^+_i a_j D\right) \equiv  \left< a^+_ia_j  \right>$.
Mean-field theory does simplify  the dynamical description of many-body systems by reducing
significantly the number of degrees of freedom to follow in time.

In nuclear physics, mean-field is often adequate to describe
average properties of one-body observables but fails to account for fluctuations.
At the wave-function level, this corresponds to a deviation of the mean-field trajectory from
the exact dynamics \cite{Lic76,Bal95}.


In order to illustrate this effect, we denote by $\delta v_{12}$ the residual two-body interaction
defined through $\delta v_{12} = H - H_{MF}$. In the weak coupling regime,
this deviation can be treated in perturbation theory and the state at time $t'$ reads \cite{Rei92}
\begin{widetext}
\begin{eqnarray}
\left| \Psi (t') \right> &=& \left| \Phi (t') \right>
-\frac{ i }{\hbar }
\int \delta v_{12}(s) \left| \Phi (s) \right> ds -\frac{ 1 }{2 \hbar^2 } T\left( \int \int \delta v_{12}(s') \delta v_{12}(s)  ds' ds
\left| \Phi (s) \right>\right).
\label{eq:pert}
\end{eqnarray}
\end{widetext}
$\delta v_{12}(s)$ corresponds to the residual interaction written in the
interaction picture,
$\delta v_{12}(s) = U^+_{MF}(s,t)\delta v_{12}(t)U_{MF}(s,t)$. While
at the mean-field level, $\left| \Psi (t') \right>$ is replaced by $ \left| \Phi (t') \right>$, due to the accumulated
effect of $\delta v_{12}$ in time, it is expected that the exact state becomes a more and more complex superposition
of a large number of Slater determinants. Accordingly, the information on the system cannot  be reduced anymore to
the knowledge of the one-body density matrix only and higher order correlations must be accounted for.
A natural extension of mean-field theory is to enlarge the number of degrees of freedom considered. This is done
for instance in the Time-Dependent Density Matrix theory (TDDM) where both the one-body density matrix and
the two-body correlation operator are followed in time \cite{Cas90}. These theories are however rarely applied due to the
large number of degrees of freedom to consider \cite{Toh02}.

An alternative way to account for correlations beyond mean-field is to use stochastic methods.
Several strategies have been proposed to introduce noise on top of the mean-field, either based on
statistical ensemble of one-body densities \cite{Ayi80,Ayi88,Ayi01}, random two-body interactions or phase shift
\cite{Bal81,Gra81}, or directly from Fermi golden rule \cite{Rei92,Rei92-2,Oni95}.
The goal in all cases, is to simulate the dynamics of correlated system by averaging over an ensemble of mean-field
trajectories. The great advantage in that case, is that the number of degrees of freedom to follow along each path
is not increased compared to standard mean-field. However, these theories are rather complex and
methods for numerical implementations are still missing.

The aim of the present work is to discuss again the possibility to replace correlated dynamics
by quantum jumps in the Hilbert space of Slater determinants. We use equation
(\ref{eq:pert}) as a starting point and we restrict the discussion to the Markovian limit.
In order to illustrate this hypothesis, we follow ref. \cite{Bal81,Gra81}.  
We assume that the residual interaction
induces random transitions treated as a statistical ensemble of two-body
interactions acting on top of the mean-field. Eq. (\ref{eq:pert}) is then replaced by a set
of evolutions with the same initial state and mean-field
but with different residual interactions.
We assume that the two-body operator has a gaussian
distribution with a mean value $\overline{ \delta v_{12}}=0$
and a second moment denoted by $\overline{ \delta v^2_{12}}$. Here,  
the average is taken over different values of $\delta v_{12}$.
In nuclear systems, the residual interaction is expected to induce transitions on a shorter time scale
(called correlation time and denoted by $\tau$) than the time associated to the mean field evolution (denoted by $\tau_{rel}$)\cite{Wei80,Lac04}.
$\tau$ is related to the average autocorrelation function
$\overline{ \delta v_{12}(s')\delta v_{12}(s)}$ which is approximated by \cite{Bal81,Gra81}
\begin{eqnarray}
\overline{ \delta v_{12}(s')\delta v_{12}(s)} \propto \overline{ \delta v^2_{12}(s)} e^{-(s-s')^2/2\tau^2}.
\label{eq:short}
\end{eqnarray}  
Using this approximation, we consider a time-scale $\Delta t$
much larger than the time $\tau$ but smaller
than $\tau_{rel}$. In the following, this limit will be called "Markovian" or "short memory time" approximation.

Using approximation (\ref{eq:short}), we consider in the following two limits for which the wave function evolution  as given
by (\ref{eq:pert}) can be replaced by quantum jumps between Slater determinants. The first case is a simplified
scenario where part of quantum correlations between $\delta v_{12}$ and $\rho$ are neglected along the path.
In this case, it is shown that the evolution can be formulated in terms of quantum jumps between many-body densities formed of
pairs of Slater determinants. In a second part, we show that the perturbative dynamics can still
be transformed into a quantum jump process even if quantal correlations are not neglected. In both cases,
all equations necessary for applications are given explicitly.

\section{fluctuations beyond mean-field in a simplified stochastic scenario}

Let us first consider the perturbative evolution of an initial Slater determinant. Assuming equation (\ref{eq:short})
and the short memory time approximation, the mean-field
does not change over $\Delta t$. Then, the average evolution of the
state, denoted by
$\overline{ \Delta \left| \Psi  \right>} = \overline{ \left| \Psi(t+\Delta t)\right>}
-\left| \Phi (t) \right>$, reduces to
\begin{eqnarray}
\overline{ \Delta \left| \Psi  \right>} = \frac{ \Delta t }{i\hbar } H_{MF} \left| \Phi (t) \right>
-\frac{ \tau \Delta t  }{2 \hbar^2} ~\overline{ \delta v^2_{12}} \left| \Phi (t) \right>.
\label{eq:average}
\end{eqnarray}      
This expression can also be regarded as an average over Markovian stochastic processes in many-body wave-functions space.
To give a deeper insight
we define the ensemble of antisymmetrized two-body residual interactions as
\begin{eqnarray}
\delta v_{12} ({\bf \sigma}) = \frac{ 1 }{4 } \sum_{\alpha \beta \gamma \delta  }a_{\alpha }^{+}a_{\beta }^{+}
\left< \alpha \beta \left| \delta v_{12} ({\bf \sigma}) \right| \delta \gamma   \right>
a_{\gamma  }a_{\delta}.
\label{eq:dv12}
\end{eqnarray}
Here, ${\bf \sigma}\equiv \left\{ \sigma_i \right\}_{i=1,N}$ where all components are stochastic variables sampled
according to gaussian probabilities with
mean zero and $\overline{{\sigma_i}^2}=1$. The number $N$ of stochastic components defines the complexity of the
process. The definition (\ref{eq:dv12}) includes the force proposed in ref. \cite{Gra81}.
Eq. (\ref{eq:average}) can be interpreted as the average over the quantum
diffusion
\begin{eqnarray}
 \Delta \left| \Psi  \right> = \left\{ \frac{ \Delta t }{i\hbar } H_{MF} + \Delta B \delta v_{12}
+ \frac{ 1 }{ 2 }\left( \Delta B  \delta v_{12} \right)^2  \right\} \left| \Phi (t) \right>,
\label{eq:ddd}
\end{eqnarray}      
where $\Delta B = i \sqrt{\tau \Delta t} /\hbar$. In the
following, we will consider last expression as a differential stochastic equation in Hilbert space
\cite{Gar00,Bre02}. We use the notation $dB$ instead of $\Delta B$ and use the Ito rules
of stochastic calculus \cite{Gar85}.
Due to the two-body nature of $\delta v_{12}$, eq. (\ref{eq:ddd}) induces complex reorganization of single-particle
degrees of freedom. After the jump, the state is not a priori a single Slater determinant.
For applications, it is highly desirable to preserve
the simple initial form of the state along the stochastic path.
This could be achieved by invoking additional
approximations described below.
Following ref. \cite{Car01,Jul02,Lac05}, we consider an initial density
\begin{eqnarray}
D = \frac{ \left| \Phi_a  \right>\left< \Phi_b  \right| }{ \left< \Phi_b \left.  \right|\Phi_a   \right>},
\label{dab}
\end{eqnarray}
where $\left| \Phi_a  \right>= {\cal A}\left( \Pi_i \left| \alpha_i  \right> \right) $ and
$\left| \Phi_b  \right> = {\cal A} \left( \Pi_i \left| \beta_i \right> \right)$
are two non-orthogonal Slater determinants formed
of products of single particle wave packets denoted respectively by $\left| \alpha_i  \right>$ and $\left| \beta_i  \right>$.
The notation ${\cal A}(.)$ corresponds to the antisymmetrized product.
We assume that both states follow the diffusion process described
by eq. (\ref{eq:ddd}) but with two independent sets of gaussian stochastic variables,
denoted respectively by ${\bf \sigma_a}$ and ${\bf \sigma_b}$. This case will be referred in the following as
the "uncorrelated noise". The use of different sets of stochastic
variables is at variance with standard quantum Monte-Carlo procedures that simulate
density evolution given by Lindblad equations\cite{Bre02}. However, this assumption has been shown
to be crucial in order to describe the exact dynamics of interacting systems with stochastic methods
\cite{Car01,Jul02}.

\subsection{Approximate stochastic mechanics in one-body density matrix space}

To approximate the diffusion process,  
we first focus on single-particle degrees of freedom.
Under the approximation
\begin{eqnarray}
\left< a^+_i a_j \delta v^2_{12} \right> && \simeq
\left< a^+_i a_j \right> \left< \delta v^2_{12} \right> + 2
\left< a^+_i a_j \delta v_{12} \right> \left< \delta v_{12} \right> \nonumber \\
&& -2\left< a^+_i a_j \right> \left< \delta v_{12}\right>^2,
\label{eq:gaus1}
\end{eqnarray}  
the one-body density evolution (eq. (\ref{eq:ddd})) reduces to
\begin{eqnarray}
d\left< a^+_i a_j \right> &\simeq &
\frac{ dt }{ i\hbar} \left< \left[ a^+_ia_j , H_{MF} \right] \right>  \nonumber \\
&+& dB_a \left( \left<  a^+_i a_j \delta v_{12} \right> - \left< a^+_i a_j\right>\left< \delta v_{12} \right>\right)
\nonumber \\
&+& dB^*_b  \left(  \left< \delta v_{12} a^+_i a_j \right> -
\left< a^+_i a_j\right>\left< \delta v_{12} \right> \right).
\label{eq:apa}
\end{eqnarray}
It is interesting to notice that, although we consider a second-order perturbation theory for the residual interaction, the second order
term exactly cancels out when approximation (\ref{eq:gaus1}) is used. Eq.
(\ref{eq:gaus1}) corresponds to a gaussian approximation
for quantal fluctuations.
Therefore, eq. (\ref{eq:apa}) provides the stochastic equation of motion of one-body degrees of freedom associated to eq.
(\ref{eq:ddd})
when neglecting part of the quantal fluctuations.
The corresponding stochastic evolution of $\rho$ reads  
\begin{eqnarray}
d \rho = \frac{ dt }{ i\hbar} \left[ h_{MF},\rho \right] &+& dB_a  (1 - \rho) U(\rho, {\bf \sigma_a}) \rho  \nonumber \\
&+& dB^*_b \rho U'(\rho, {\bf \sigma_b}) (1 - \rho),  
\label{eq:drho}
\end{eqnarray}
where $h_{MF}$ denotes the matrix elements associated to the mean-field Hamiltonian  
while $U(\rho, {\bf \sigma_a})=Tr_2(\delta v_{12} ({\bf \sigma_a}) \rho_2)$ and $U'(\rho, {\bf \sigma_b}) = Tr_2(\rho_2 \delta v_{12}( {\bf \sigma_b}))$.

The stochastic one-body evolution given by eq. (\ref{eq:drho}) contains also part of the information on correlations.
Indeed, an approximate evolution of the two-body density, whose matrix
elements are $\left<  a^+_i a^+_j a_l a_k \right>= \left<  kl \left| \rho_{12} \right|ij \right>$
can be obtained through approximations similar to eq. (\ref{eq:gaus1}) but preserving the symmetry of the two-body density:
\begin{widetext}
\begin{eqnarray}
< a^+_i a^+_j  a_l a_k v_{12} >  &\simeq &
< a^+_i a^+_ja_l a_k> <v_{12}>  
+ \left( <a^+_ia_k v_{12}> <a^+_j a_l > -  <a^+_ia_l v_{12}> <a^+_j a_k > \right) \nonumber \\
&+&   \left( <a^+_i a_k > <a^+_j a_l v_{12} > -  <a^+_i a_l > <a^+_j a_k v_{12} > \right)
\nonumber \\
&-&
 2 \left( <a^+_ia_k><a^+_j a_l> -  <a^+_ia_l> <a^+_j a_k>  \right)< v_{12}>
\label{eq:gaus11}
\end{eqnarray}
and
\begin{eqnarray}
< a^+_i a^+_j  a_l a_k v_{12} v_{12} >
&\simeq&  
< a^+_i a^+_j a_l a_k> <v_{12}^2> +
 2 \left( <a^+_ia_k v_{12}> <a^+_j a_l v_{12}> -
<a^+_ia_l v_{12}> <a^+_j a_k v_{12}> \right)\nonumber \\
&-& 2 \left( <a^+_ia_k><a^+_j a_l> -  <a^+_ia_l><a^+_j a_k> \right)< v_{12}>^2.
\label{eq:gaus12}
\end{eqnarray}
\end{widetext}
Combining with Ito rules,
the evolution of $\rho_{12}$ reduces to
\begin{eqnarray}
d< a^+_i a^+_ja_l a_k> \simeq
d (\rho_{ki}\rho_{lj} - \rho_{kj}\rho_{li})
\label{eq:sep}
\end{eqnarray}
indicating that the two-body evolution can be deduced from the stochastic evolution of $\rho$.
Although eq. (\ref{eq:sep}) is similar to the mean-field case, it contains
correlations beyond mean-field.  
A similar situation occurs in the exact reformulation of self-interacting fermions with quantum jumps \cite{Lac05}.

In summary, the jump process described by eq. (\ref{eq:ddd}) for both state vectors entering in $D$
can be approximated by the jump process in one-body space given by eq. (\ref{eq:drho})
if part of the quantal fluctuations are neglected. The advantage of this approximation is that
expression (\ref{dab}) for $D$ is preserved along the stochastic path. In this work, we restrict ourself to this
limit and eq. (\ref{eq:drho}) will be referred to the incoherent Stochastic Mean-Field (SMF) dynamics.
The properties of this diffusion process are described below.  

We consider that the single-particle states of $\left| \Phi_a  \right>$ and $\left| \Phi_b  \right>$   initially verify
\begin{eqnarray}
\left<  \beta_j \left.  \right| \alpha_i \right>= \delta_{ij}.
\label{eq:ortho}
\end{eqnarray}
Eq. (\ref{eq:drho}) can be simulated by quantum jumps for single-particle states given by
\begin{eqnarray}
\left\{
\begin{array}{ccc}
\left| {d \alpha_i} \right> &=& \left[ \frac{dt}
{i \hbar} h_{MF}(\rho) + dB_a (1- \rho) U(\rho, {\bf \sigma_a}) \right]
\left| {\alpha_i} \right> \\
&& \\
\left< d \beta_j \right| &=& \left< \beta_{j} \right| \left[ -\frac{dt}
{i \hbar} h_{MF}(\rho) + dB^*_b  U'(\rho,{\bf \sigma_b}) (1 - \rho)\right].
\end{array}
\right.
\end{eqnarray}
The latter quantum diffusion process has several attractive aspects.
First, it can be easily verified that eq. (\ref{eq:ortho}) is preserved along the stochastic path.
Thus, the one-body density reads at all time $\rho =
\sum_i \left|  \alpha_i \right>\left<  \beta_i \right|$. Consequently,
the trace of the density is constant
along the path: $Tr(dD) = Tr(d\rho)=0$. In addition, $\rho$ remains a projector, i.e.
$\rho^2 = \rho$ at all time. Finally, the total entropy $S=-k_B Tr(D \ln D)$ is constant
along the path. Indeed, since the density is given by eq. (\ref{dab}), $S(D)$
identifies with the one-particle entropy $S(\rho)$.
Using  equation (\ref{eq:drho}) and Ito rules, we obtain $\overline{dS(\rho)}=0$.
Despite a constant entropy, the SMF induces correlations
beyond mean-field.
Indeed, starting from an initial two-body density $\rho_{12} = {\cal A}(\rho_{1}\rho_{2})$, after one time step,
the
average evolutions of the one- and two-body density matrices read
\begin{eqnarray}
\left\{
\begin{array} {lll}
d\overline{\rho} &=& \frac{ dt }{i \hbar } \left[ h_{MF}, \rho \right] \\
\\
d\overline{\rho_{12}} &=& \frac{ dt }{i \hbar } \left[ h_{MF} (1) + h_{MF} (2), \rho_{12} \right] + dC_{12}. \\
\end{array}
\right.
\end{eqnarray}
The labels $"1"$ and $"2"$ refer to the particle on which the operator is acting \cite{Abe96}.
$dC_{12}$ corresponds to correlations beyond mean-field associated to the
stochastic one-body evolution given by eq. (\ref{eq:drho}). It reads
\begin{eqnarray}
dC_{12} = -\frac{ \tau dt }{ \hbar^2} && \left\{ (1 - \rho_1)(1  - \rho_2) \overline{U_1({\bf \sigma_a}) U_2({\bf \sigma_a})} \rho_{12} \right. \nonumber \\
&& \left. + \rho_{12}\overline{U'_1({\bf \sigma_b}) U'_2({\bf \sigma_b})}(1 - \rho_1)(1 - \rho_2) \right\},
\label{eq:dcor}
\end{eqnarray}
where the density dependence are omitted in $U$ and $U'$.
Eq. (\ref{eq:dcor}) clearly indicates
 that $dC_{12}$ is a second order
term in perturbation. Note that it has a similar form as the second
moment of the initial stochastic correlation used in ref. \cite{Ayi01}.

\begin{figure}[tbph]
\includegraphics[height=7.cm]{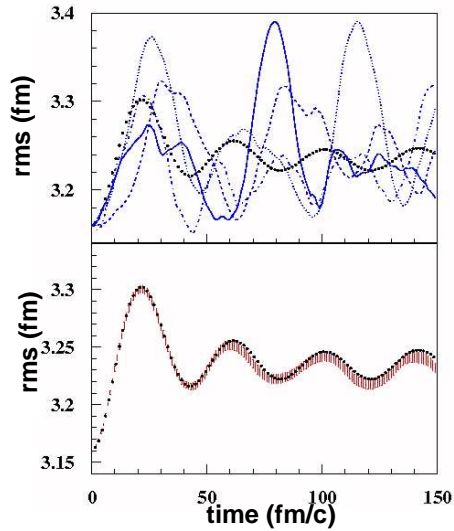}
\caption{Top: Evolution of the root mean square radius (rms) as a function of time. Black circles correspond
to the standard TDHF evolution while different lines correspond to different stochastic paths.
Bottom: Error bars correspond to the rms evolution obtained by averaging over different paths while black
circles correspond to the TDHF case. The stochastic simulation is performed for $g_0=500$ MeV/fm.
The average is taken over 200 trajectories. The width of the error bars correspond to the statistical fluctuations of the rms.}
\label{fig:esmf1}
\end{figure}

\subsection{Illustration of application}
\label{secapsto}
To illustrate the SMF theory, we consider the monopolar vibration of a $^{40}$Ca nucleus.
The system is initially prepared in a pure state $D=\left| \Phi  \right>
\left< \Phi  \right|$, where $\left| \Phi  \right>$ is a Slater determinant
solution of a constrained
Hartree-Fock (CHF) equation.
The CHF equation is solved assuming spherical symmetry
and spin and isospin saturation.  
The Skyrme interaction of ref. \cite{Str79} is used
in the mean-field.
We assume in addition to the self-consistent mean-field,
a monopolar constraint $\lambda r^2$ with $\lambda = 0.25$ MeV.fm$^{-2}$  at $t<0$ fm/c \cite{Gia88}.
At $t=0$ fm/c, the constraint is relaxed and two dynamical calculations are considered. The
first one corresponds to the time-dependent Hartree-Fock (TDHF) evolution.
In the second case, the SMF evolution described by eq. (\ref{eq:drho}) is performed with
a statistical ensemble of contact interactions
defined by one stochastic variable, i.e.
$\delta v_{12}(\sigma) = \sigma v_{res}$ where  $v_{res}$ is a contact interaction. In this case,
$U(\sigma,r)$ takes the form
$U(\sigma,r) =  \sigma g_0 \rho(r)$, where $g_0$ is a parameter measuring the strength of the perturbation.
In both cases, evolutions are solved assuming spherical symmetry.

\begin{figure}
\begin{center}
\includegraphics[height=7.cm,angle=-90.] {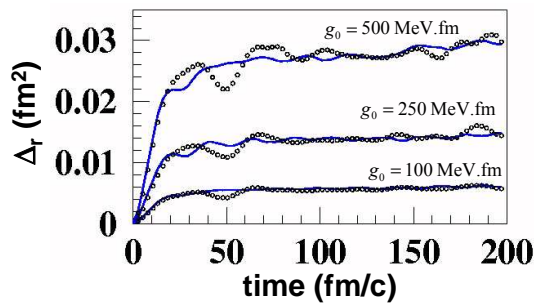}
\end{center}
\caption{Evolution of the dispersion of the rms as a function of time for different values of $g_0$. The different curves
from bottom to top corresponds respectively to $g_0=100$, $250$ and $500$ MeV.fm.
Solid lines and open circles correspond
respectively to an initial constraint $\lambda = 0$ MeV.fm$^{-2}$  and $\lambda = 0.25$ MeV.fm$^{-2}$.
}
\label{fig:esmf2}
\end{figure}

The evolution of the root mean square radius (rms)
obtained with TDHF is presented in figure \ref{fig:esmf1} (filled circles).
The different lines displayed on the top part of figure \ref{fig:esmf1} correspond to the
evolution of the rms along several
stochastic paths obtained with $g_0=500$ MeV/fm and a collision time $\tau = 0.01$ fm.
In each case, the stochastic evolution differs
significantly from the mean-field prediction. Bottom part of
figure \ref{fig:esmf1} shows a comparison between the TDHF evolution and the
evolution of the rms obtained by averaging over the different stochastic trajectories.
Interestingly enough, the average evolution identifies with the TDHF evolution. This example illustrates
a special situation where the mean-field dynamics can be recovered from complex trajectories in many-body
space \cite{Zel93}. However, significant fluctuations around the mean TDHF trajectories are observed. This is
illustrated in figure \ref{fig:esmf2} where the quantity
$\Delta_r = \sqrt{\overline{\left< r^2 \right>^2} -\overline{\left< r^2 \right>}^2}$
is displayed as a function of time for different values of $g_0$ and $\lambda$.
We also computed as a reference the mean-field width $\sigma_{MF}$, corresponding to the quantal fluctuations of $r^2$
estimated in the mean-field approximation, leading to $\sigma_{MF} = 6.3 fm^2$. As we do expect in the weak coupling
approximation, the additional fluctuations $\Delta_r$ induced by the stochastic term are much smaller than $\sigma_{MF}$.    
It turns out that the dispersion is properly parametrized by the formula $\Delta_r =\Delta_0 (1-e^{-\Gamma_0 t})$ where
$\Delta_0$ is proportional to $g_0$ while $\Gamma_0$ is independent of it.    
Therefore, while the average evolution of the rms collective variables is not affected by the stochastic process,
fluctuations around the mean value increases and saturates as we do expect in Brownian motion. Interestingly
enough, the behavior observed here is very similar to the description of a quantum oscillator \cite{Rug82} with Nelson
stochastic mechanics \cite{Nel66} replacing $(\hbar/m)^{1/2}$ by $g_0$. Assuming a single collective state
and using similar techniques as in ref. \cite{Rug82}, analytical expression can be obtained for $\Delta_r$
where $\Delta_0$ is indeed proportional to $g_0$ while $\Gamma_0$ depends only on the oscillator frequency.
Note however that a complete understanding of the
Brownian process presented here pass through the linearization
of eq. (\ref{eq:drho}) as in ref. \cite{Lac04}.

\subsection{Critical discussion}

In this section, it has been  shown that the presented SMF theory can be applied to account for fluctuations beyond
mean-field. The above example can only serve as an illustration due to the very schematic residual
interaction used and to the very small time $\tau$ (see estimation in ref. \cite{Ayi94,Kol95}).
Besides the simplicity of the force, it is important to note that the above theory
only gives a partial answer to the simulation of correlations beyond mean-field for realistic nuclear systems.
Indeed, as clearly seen in figure (\ref{fig:esmf1}), while fluctuations of one-body observables are increased,
the average evolution of the rms identifies with the TDHF case. Therefore, no additional damping is observed  
in SMF in opposite to what is obtained in extended Time-Dependent Hartree-Fock (extended TDHF) \cite{Lac04}.
Indeed, dissipative aspects present in
the memory kernel of extended TDHF  are not included in the present framework.
A careful analysis demonstrates that
the absence of a collision term could be assigned to the approximations made on quantal fluctuations
(equation (\ref{eq:gaus1}), (\ref{eq:gaus11}) and (\ref{eq:gaus12})).

Under these approximations, it is however possible to show that a fermionic system submitted to a statistical
ensemble of residual interactions $\delta v_{12}$ can be treated by a jump process related to its mean-field
(equation (\ref{eq:drho})). The form of the noise is a second critical aspect for nuclear physics.
In fact, it is expected that the residual interaction is dominated by the two-particle two-holes (2p-2h) channels.
Due to the mean-field nature of the noise in eq. (\ref{eq:drho}), these components cancel out in the stochastic part.
Therefore, while the above SMF can be of great interest, some important aspects for nuclear physics are missing. In the next section, 
we discuss the possibility to
use quantum jump process in one-body space in a more general framework.

\section{Stochastic mechanics with dissipation}

In order to generalize the stochastic method described previously, we consider directly the evolution of the
many-body density $D$ associated to eq. (\ref{eq:pert}). Using the short memory time
approximation for $\delta v_{12}$, the evolution of $D$ can be recast as
\begin{eqnarray}
\Delta D = \frac{ \Delta t}{i \hbar} [H_{MF},D] -\frac{\tau \Delta t}{2\hbar^2}
\overline {\left[\delta v_{12}, \left[\delta v_{12},D \right] \right]}.
\label{eq:taud}
\end{eqnarray}
In the previous section, the term $\overline{\delta v_{12} D \delta v_{12}}$ has been neglected due to the fact that the
average evolution of wave-functions were directly considered. Due to this approximation, it was possible to have
uncorrelated noise for $\left| \Phi_a \right>$ and $\left< \Phi_b \right|$. Here, this contribution
is not neglected. As a consequence, the uncorrelated noise assumption is not possible anymore.

The one-body density matrix
equation of motion associated to (\ref{eq:taud}) is given by  
\begin{eqnarray}
\frac{d \rho}{dt} &=& \frac{1}{i\hbar}\left[ h_{MF}(\rho),\rho \right] -  \frac{g}{2} {\cal D}(\rho),
\label{eq:drho2}
\end{eqnarray}
where $g$ is a real constant, $g = \tau/\hbar^2$.
${\cal D}(\rho)$, called "dissipator" hereafter, corresponds to the average effect of the
residual interaction and reads
\begin{eqnarray}
\left< j \left| {\cal D} \right| i \right> &=& \overline{
\left<  \left[\left[a^+_i a_j , \delta v_{12}  \right], \delta v_{12} \right]  \right>}.
\label{eq:diss}
\end{eqnarray}
We assume that the system is initially in a pure state described by a Slater determinant $\left| \Phi (t) \right>$
formed of $N$ orthonormal single particle states denoted by $\left| \alpha  \right>$.
The associated initial one-body density matrix reads $\rho = \sum_{\alpha } \left| \alpha  \right>
\left< \alpha  \right|$. Having in mind the nuclear many-body problem, we assume that only 2p-2h components of
$\delta v_{12}$ are not equal to zero. Completing the
hole states by a set of particle states, denoted by $\left| \tilde{\alpha} \right>$, we have:    
\begin{eqnarray}
\delta v_{12} (t) = \frac{ 1 }{4 } \sum_{\tilde{\alpha} \tilde{\beta} \alpha \beta}
a_{\tilde{\alpha}}^{+}a_{\tilde{\beta}}^{+}
\left< \tilde{\alpha} \tilde{\beta} \left| v_{12}(t) \right| \alpha \beta   \right>
a_{\alpha }a_{\beta}.
\label{eq:dv12ph}
\end{eqnarray}
${\cal D}$ can then be recast as
\begin{eqnarray}
{\cal D}(\rho) =  Tr_2 \left[ v_{12}, F_{12} \right],
\label{eq:dissrho}
\end{eqnarray}
where $F_{12}$ is equal to  
\begin{eqnarray}
F_{12} &=& \frac{1}{2} \left\{ \left( 1 - \rho_1\right)\left( 1
-\rho_2  \right)v_{12} \rho_1 \rho_2 \right.  \nonumber \\
&& - \left. \rho_1 \rho_2 v_{12} \left( 1 - \rho_1\right) \left( 1- \rho_2  \right) \right\}.
\label{eq:F12}
\end{eqnarray}
Expression (\ref{eq:dissrho}) takes a form similar to the
collision term generally obtained in extended TDHF \cite{Lac04}. The dissipator
${\cal D}(\rho)$ can be further transformed. Indeed, $\delta v_{12}$ given by eq. (\ref{eq:dv12ph})
can always be decomposed as (see for instance \cite{Jul01})
\begin{eqnarray}
\delta v_{12} = -\frac{1}{4} \sum_n \lambda_n {\cal O}^2_n,
\label{eq:dvoo}
\end{eqnarray}  
where $\lambda_n$ are real and the ${\cal O}_n$ correspond to a set of commuting Hermitian one-body operators written as
${\cal O}_n = \sum_{\tilde{\alpha}\alpha} \left< \tilde{\alpha} \left| O_n \right| \alpha \right>a^+_{\tilde{\alpha}} a_\alpha$.
Reporting in eq. (\ref{eq:dissrho}), ${\cal D}(\rho)$ can be recast as
\begin{eqnarray}
{\cal D}(\rho)  = \sum_{mn} \Gamma_{mn} \left[O_n O_m \rho +  \rho O_n O_m-  2 O_m \rho O_n
\right].
\label{eq:disslast}
\end{eqnarray}
The coefficient $ \Gamma_{mn}$ are given by  
\begin{eqnarray}
\Gamma_{mn} = \frac{1}{2} \lambda_m \lambda_n Tr(O_m (1-\rho) O_n \rho).
\end{eqnarray}
We recognize in this expression, the quantum covariance between the operator ${\cal O}_n$ and ${\cal O}_m$ , i.e.
$Tr(O_m (1-\rho) O_n \rho) = \left< {\cal O}_m {\cal O}_n \right> - \left< {\cal O}_m\right>\left< {\cal O}_n\right>$.
Expression (\ref{eq:disslast}) has the form of the dissipator appearing usually in the Lindblad equation\cite{Bre02}.
Therefore, we have shown that the evolution of one-body degrees of freedom associated to equation (\ref{eq:taud})
identifies with a Markovian quantum master equation generally obtained in quantum open systems. A large amount of work
is devoted to the simulation of such master equation by quantum jump methods (see for instance
\cite{Dio86,Car93,Rig96,Ple98,Bre02}) and one can take advantage of the most recent advances in this field.
This aspect has however rarely been discussed in the context of self-interacting system. In the following, 
the associated diffusion process is precised and we show that it indeed corresponds to jumps between Slater determinants.
The stochastic Schroedinger equation for single-particle wave-function is finally given.

\subsection{Explicit form of the stochastic process}

Following ref. \cite{Bre02}, we introduce the Hermitian positive matrix $\Gamma$ with components $\Gamma_{mn}$.
An economical method to introduce quantum jump process \cite{Bre02}
is to use the unitary transformation $u$ that diagonalizes $\Gamma$, i.e. $\Gamma  = u^{-1}\gamma u$,
where $\gamma$ is the diagonal matrix of the eigenvalues of $\Gamma$.
New operators $A_k$ can be defined
by the transformation $A_k = \sum_n u^{-1}_{k n} O_n$. The dissipator is then recast as  
\begin{eqnarray}
{\cal D}(\rho)  = \sum_{k} \gamma_{k} \left[A^2_k \rho +  \rho A^2_k -  2 A_k \rho A_k
\right].
\label{eq:dissverylast}
\end{eqnarray}
The last equation can be simulated using the average over the stochastic mean-field dynamics:
\begin{eqnarray}
d \rho &=& \frac{dt}{i\hbar}\left[ h_{MF}(\rho),\rho \right] -  g\frac{ dt }{2} {\cal D}(\rho) + db_{sto},
\label{eq:drhosto}
\end{eqnarray}
where $db_{sto}$ is a stochastic one-body operator which,  using Ito rules \cite{Gar85}, reads
\begin{eqnarray}
db_{sto} &=&\sum_k  \left\{ dW_k (1 - \rho)A_k \rho + dW^*_k \rho A_k (1 - \rho) \right\}.
\end{eqnarray}
Here $dW_k$ denotes stochastic variables given by $dW_k  = -i d\xi_k \sqrt{g \gamma_k}$,
where $d \xi_k$ corresponds to a set of real gaussian stochastic variables with mean zero and
$d\xi_k d\xi_{k'} = \delta_{kk'}dt$.

\subsection{Nature of the stochastic process in Hilbert space}
It is worth noticing that the proposed dissipative equation and its stochastic counterpart are
only well defined if the density is initially prepared as a pure Slater-determinant state. We now turn to
the essential properties of equation (\ref{eq:drhosto}). First, it preserves the number of particles $Tr(d\rho) = 0$.
In addition, if initially $\rho^2 = \rho$, then
\begin{eqnarray}
d\rho d\rho -g\frac{dt}{2} \left[\rho {\cal D}(\rho) + {\cal D}(\rho)\rho  \right] = -g\frac{dt}{2} {\cal D}(\rho)
\end{eqnarray}
which is obtained using Ito stochastic rules and retaining only terms linear in $dt$. The last expression demonstrates
that $(\rho+d\rho)^2 = \rho+ d\rho$. Thus, $\rho$ remains a projector along the stochastic path. As a consequence,
the pure state nature of the many-body density matrix is preserved along the stochastic
path, i.e. $D=\left| \Phi(t)  \right> \left< \Phi (t) \right|$ where $\left| \Phi  \right>$ is a normalized
Slater determinant at all time. The associated stochastic Schroedinger equation for single-particle states reads
\begin{widetext}
\begin{eqnarray}
d\left| \alpha  \right> = \left\{ \frac{dt}{i\hbar} h_{MF}(\rho) + \sum_k dW_k (1-\rho) A_k
-  g\frac{dt}{2} \sum_k \gamma_k \left[ A^2_k \rho + \rho A_k \rho A_k -2 A_k \rho A_k \right]
   \right\}
\left| \alpha  \right>.
\label{eq:dsto1}
\end{eqnarray}
\end{widetext}
This last expression can be directly used for practical applications.
 
In this section, assuming that the residual interaction can be written in terms of $2p-2h$ components, we have shown that
the dissipative dynamics of one-body degrees of freedom can be simulated by quantum jumps.
The stochastic method differs significantly from the simplified scenario considered in previous
section. First, the above theory does not require to introduce generalized many-body density matrix formed of two
Slater determinants since $D=\left| \Phi(t)  \right> \left< \Phi (t) \right|$ along each path. As a counterpart, the
numerical effort required to treat the new SMF which includes dissipative aspects is significantly increased.
Indeed, in section \ref{secapsto}, we have shown that fluctuations can be introduced
using simplified residual interaction and a limited number of noise terms. In the new SMF, the numerical effort is
directly proportional to the number of one-body operators entering in eq. (\ref{eq:dvoo}).
This situation is similar to stochastic methods used in nuclear structure studies \cite{Koo97}. In that case,
large numerical effort is required.
For instance, if we assume that a physical system is described in a mesh, then the number of noise terms is
is a priori as large as the number of mesh points.
The numerical implementation of the quantum jump process in its full complexity is expected to be still difficult with
present computers capacities.
Therefore, specific numerical techniques as well as truncation procedures should be developed in order to implement the method
in many-body dissipative dynamics.

\section{Conclusion}

In this work, we have presented a discussion on the possibility to replace the dynamics
of interacting fermions by quantum jumps. Assuming a weak coupling approximation
and a short memory time for the residual interaction, two scenarios have been considered.
Focusing on one-body degrees of freedom, two different approximations lead to equations
of motions for the one-body density matrix that can be treated by quantum jumps between Slater determinants.

In the first stochastic mean-field process, part of the quantal correlations between $\rho$ and $\delta v_{12}$
are neglected along the path. The SMF is illustrated in the monopolar vibration
of a calcium nucleus. In this case, while expectation values of one-body observables are unchanged,
fluctuations are increased compared to mean-field.
In the presented application, the residual interaction is rather simple.
However, more complex statistical ensembles can be used like two-body random
interactions. This method suffers from the absence of dissipative effects and appears
to be too simplified for the  nuclear many-body problem.

In the second part of this work, we show that the dynamics of correlated systems can be
described by a quantum jump process even if no approximation on the quantal correlations
are made. Restricting to 2p-2h residual interactions, a guideline is given for transforming
the evolution of the one-body density matrix evolution into a Lindblad equation.
The stochastic process corresponds to
quantum jumps between pure state densities $D = \left| \Phi  \right>\left< \Phi  \right|$
where $\left| \Phi \right>$ is a Slater determinant. The associated stochastic equation of motion
for single-particle wave-functions is given.  

Finally, we would like to mention that the presented framework does not account for non-Markovian effects. It has however
been shown in ref. \cite{Lac98-2} that the memory effect might be important in the nuclear context. Promising
work are actually devoted to incorporate non-Markovian effects in quantum Monte-Carlo
methods \cite{Dio96,Dio98,Str99,Bre99}.

{\bf ACKNOWLEDGMENTS}

The author is grateful to Thomas Duguet, Dominique Durand and Cedric Simenel for
the careful reading of the manuscript and to Olivier Juillet for
discussions during this work.




\end{document}